%% file: Template.tex
\documentclass{article}
\usepackage{spconf,amsmath,graphicx}
\usepackage{amsmath,dsfont,bbm,epsfig,amssymb,amsfonts,amstext,verbatim,amsopn,cite,subfigure,multirow,multicol,lipsum,xfrac}
\usepackage{amsthm}
\usepackage{mathtools,amsthm}
\usepackage{perpage}
\usepackage{balance}
\usepackage{url}
\usepackage{amsfonts}
\usepackage{epsfig}
\usepackage[font={small}]{caption}

\usepackage[letterpaper,
left=.75in, right=.75in,
top=1in, bottom=1in
]{geometry}

\usepackage{psfrag}	
\usepackage{etoolbox}
\usepackage{algorithmicx}
\usepackage[Algorithm,ruled]{algorithm}
\usepackage{algpseudocode}
\usepackage{pifont}
\usepackage[utf8]{inputenc}
\usepackage[T1]{fontenc}  
\usepackage[nolist]{acronym}
\MakePerPage{footnote}
\usepackage{paralist}
\usepackage{enumitem}
\usepackage{bbm}
\usepackage[process=auto]{pstool}
\usepackage{tikz}
\usetikzlibrary{shapes,arrows}

\title{Maximum-A-Posteriori Signal Recovery with Prior Information: Applications to Compressive Sensing}
\name{Ali Bereyhi and Ralf R. M\"uller\thanks{This work was supported by the German Research Foundation, Deutsche Forschungsgemeinschaft (DFG), under Grant No. MU 3735/2-1.}
}
\address{Friedrich-Alexander Universit\"at Erlangen-N\"urnberg\\ ali.bereyhi@fau.de and ralf.r.mueller@fau.de}
\include{commands}

\begin{document}
%\ninept
%
\maketitle

%%%%%%%%%%%%%%%%%%%%%%%%%%%%%%%
%%%TIKZ blocks%%%%%%%%%%%%%%%%%
\tikzstyle{block} = [draw, rounded corners, rectangle, minimum height=2.5em, minimum width=5em]
\tikzstyle{margin} = [draw, dotted, rectangle, minimum height=2.1em, minimum width=2em]
\tikzstyle{sum} = [draw, circle, node distance=1cm, inner sep=0pt]
\tikzstyle{input} = [coordinate]
\tikzstyle{output} = [coordinate]
\tikzstyle{pinstyle} = [pin edge={to-,thick,black}]
%%%%%%%%%%%%%%%%%%%%%%%%%%%%%%%%%%%%%%%%%%%%%%%%%%%%%%%%%%%%%%%%%%%%%%%%%%%%%%%%%%%

%
\begin{abstract}
This paper studies the asymptotic performance of maximum-a-posteriori estimation in the presence of prior information. The problem arises in several applications such as recovery of signals with non-uniform sparsity pattern from underdetermined measurements. With prior information, the maximum- a-posteriori estimator might have asymmetric penalty. We consider a generic form of this estimator and study its performance via the replica method. Our analyses demonstrate an asymmetric form of the decoupling property in the large-system limit. Employing our results, we further investigate the performance of weighted zero-norm minimization for recovery of a non-uniform sparse signal. Our investigations illustrate that for a given distortion, the minimum number of required measurements can be significantly reduced by choosing weighting coefficients optimally. %compared to the case with uniform weighting.
\end{abstract}
%in this case which results in asymmetry of the estimation scheme.
%\IEEEpeerreviewmaketitle

\section{Introduction}
%Consider the linear sampling system
The problem of estimating $\bx\in\setX^N$, for some $\setX\subset \setR$, from %noisy measurements % vector
\begin{align}
\by=\mA\bx+\bz, \label{eq:lin}
\end{align}
with $\mA\in\setR^{K\times N}$ and $\bz \sim \mathcal{N}(0,\lambda_0\mI_K)$, arises in various applications. In presence of prior information, the \ac{map} estimation approach might deal with an asymmetric penalty term appearing due to the non-identical prior distributions. In this paper, we intend to investigate the asymptotic performance of this class of estimators which encloses several reconstruction schemes in signal processing.

%
%In the absence of prior information, the approach is to assume the source entries identically distributed. This assumption results in the design of symmetric estimators in terms of source entries, e.g., \ac{map} estimator with a Gaussian prior. The availability of prior information, however, impacts this classical approach in the two senses.
%\begin{inparaenum}
%\item The assumption of identically distributed entries is not necessarily realistic.
%\item The designed estimator, which takes the prior information into account, may deal with asymmetric penalty.
%\end{inparaenum}
%In this paper, we are interested on investigating the impact of these asymmetries in the asymptotic regimes. %For this aim, an asymmetric form of the \ac{map} estimator is considered which encloses a large class of reconstruction schemes.~The~asymptotics of the estimator is then studied for a general statistical model of the source which addresses a wide range of scenarios with prior information available at the estimator.~Our~investigations show that for this general setup, an asymmetric version of the decoupling property holds in the large-system limit. 
%\subsection*{Motivation, Tools and Prior Studies}
%
%The motivation behind this study comes from the literature of non-uniform sparsity models and their corresponding reconstruction techniques. 
Particular examples of these estimators are the weighted norm minimization schemes \cite{candes2008enhancing} in compressive sensing \cite{donoho2006compressed,candes2006stable} which are employed for recovery of signals with non-uniform sparsity patterns. In this problem, the signal consists of multiple sparse blocks whose sparsity factors are different. A restricted class of such non-uniform sparse settings, in which the signal support is partially known, was addressed in \cite{vaswani2010modified}, and the modified-CS scheme was proposed for signal recovery. Weighted $\ell_1$-norm minimization was further invoked in \cite{khajehnejad2009weighted} for non-uniform sparse recovery in which different blocks of signal samples have different sparsity factors. More general settings were investigated in recent studies; see \cite{jacques2010short,mansour2012support,scarlett2013compressed,oymak2012recovery,bah2016sample,rauhut2016interpolation} and the references therein. 
%
%
%The investigations in \cite{vaswani2010modified} depicted that sufficient conditions for exact recovery in this case can be significantly weaker compared to that for conventional recovery schemes which ignore the known support. 
%The non-uniform model also arises in other applications. %such as multiuser communications. 
%An example is a multiuser uplink system with users transmitting statistically different signals. %due to their physically different conditions. 
%The studies in \cite{guo2005randomly,rangan2012asymptotic} addressed partially this problem by considering different transmit powers for the users.% receive symbols at the transmitter. %Nevertheless, the asymmetry of the source distribution in other respects has not been yet precisely addressed in~the~literature.
%\subsection*{The Replica Method and Decoupling Property}

Due to the nonlinear nature of the \ac{map} estimator, basic tools fail to investigate its large-system~performance.~Several studies thus invoked the replica method for investigation. This method was developed for analysis of spin glasses \cite{edwards1975theory} in the physics literature and accepted as an efficient mathematical tool in information theory; e.g., \cite{tanaka2002statistical}. The method was moreover employed to investigate the performance of various recovery schemes in large compressive sensing systems \cite{tulino2013support,vehkapera2016analysis,bereyhi2016statistical,bereyhi2017replica}. For non-uniform sparse models, the method was employed in \cite{tanaka2010optimal} to study the performance of weighted $\ell_1$-norm minimization recovery considering  noise-free measurements. In this paper, we consider a generic class of estimators which includes formerly studied schemes such as weighted $\ell_1$-norm minimization and also encloses several other settings whose performances have not yet been addressed in the literature. Invoking our results we derive an asymmetric version of the \ac{map} decoupling principle which extends the results of \cite{rangan2012asymptotic,bereyhi2016rsb} to a larger class of estimators.
\section{Problem Formulation}
\label{sec:sys}
Consider \eqref{eq:lin} with ${K}/{N}=\alpha < \infty$ as ${N \uparrow \infty}$. Let $[N]\coloneqq\set{1,\ldots,N}$ be partitioned into disjoint subsets $\setN_j$ for $j\in[J]$. $J$ is assumed to be fixed and bounded meaning that ${J}/{N} \downarrow 0$ as $N$ grows large. The signal $\bx_{N\times 1}$ is divided into $J$ blocks. The block $j$ is denoted by $\setB_j (\bx)$ and contains entries whose indices are in $\setN_j$, i.e., $\setB_j (\bx) \coloneqq \set{x_n:  n\in\setN_j }$. We use the notation $j(n)$ to denote the index of the block to which $x_n$ belongs, i.e., $x_n\in\setB_{j(n)}(\bx)$. The entries of $\bx$ are independent, and $x_n\sim\mathrm{p}_{j(n)}(x_n;\rho_{n})$ where $\set{\rho_{n}}$ is a deterministic sequence over $[N]$. %An example of such $\bx$ is a real signal with non-uniform sparsity pattern in which for $j\in[J]$ %each signal entry has different probability of being zero. In this case, the signal is 
%the cumulative distribution
%\begin{align}
%\rmp_{j(n)}(x_n;\rho_{n}) = \rho_n \rmq_{j(n)}(x_n) + (1-\rho_n) \delta (x_n) \label{eq:cs_source}
%\end{align}
%for some real distribution $\rmq_{j(n)}(x_n)$. This signal can be equivalently represented with $J$ real random variables $s_1,\ldots, s_J$ in which $s_j\sim\rmq_{j}(s_j)$ and $N$ binary random variables $b_1, \ldots,$ $b_N$ in which $\Pr\set{b_n=1}=\rho_n$. The $n$th entry of the signal then reads $x_n=s_{j(n)} b_n$.
The signal is reconstructed from $\by$ as %the \ac{map} estimator $\bgg(\cdot)$ as $\bhx=\bgg(\by)$ where
\begin{align}
\bhx =  \argmin_{\bv\in\setX^n} \frac{1}{2\lambda} \norm{\by-\mA\bv}^2+u(\bv;\bc) \label{eq:map}
\end{align}
where $\lambda$ is the estimation parameter, $\bc_{N\times1}$ contains weighting coefficients $\set{c_n}$, and $u(\bv;\bc)$ is a penalty function with decoupling property, i.e., there exist $\set{u_{j}(v_n;c_{n})}$ such that
\begin{align}
u(\bv;\bc)=\sum_{n=1}^N u_{j(n)}(v_n;c_{n}).%\\
%&=\sum_{n=1}^N u_{j(n)}(v_n;c_{n}).
\end{align}
%and its empirical distribution converges to $\rmq_c^\hj$ as $N_j\uparrow \infty$.
%\subsection{Special Cases}
%The estimator is matched to the source distribution when for all $n\in [N]$, $j\in[J]$ and $v\in\setX$, we set $c_n=\rho_n$, $\lambda=\lambda_0$ and $u_j(v;c_{n})=-\log \rmp_j (v;\rho)$. To insure the certainty of the estimation, we further assume that the objective function in \eqref{eq:map} has a unique minimizer.
%\begin{enumerate}[label=(\alph*)]
%\setcounter{enumi}{7}
%\item 
%\end{enumerate}
$\mA$ is assumed to be random, such that $\mJ=\mA^{\trp} \mA= \mU \mD \mU^{\trp}$ with $\mU$ being Haar distributed and $\mD$ denoting the diagonal matrix of eigenvalues. A trivial example is a matrix with \ac{iid} entries. The empirical distribution of eigenvalues when $N \uparrow \infty$ is denoted by $\mathrm{p}_{\mJ}(\lambda)$. For this distribution, the Stieltjes transform~is~given~by $\rmG_{\mJ}(s) = \E{(\lambda-s)^{-1}}$ where $\lambda \sim \rmp_{\mJ} (\lambda) $, and the $\rmR$-trans-form is defined as $\rmR_\mJ (\omega) \coloneqq \rmG_\mJ^{-1} (-\omega) - \omega^{-1}$ with $\rmG_\mJ^{-1} (\cdot)$ being the inverse \ac{wrt} composition.

\hspace*{-1mm}The setting recovers several problems in signal processing. An example is recovery of non-uniform sparse signals from noisy measurements in compressive sensing:~Let~$J=1$~and %$\rmp_1(x_n;\rho_n)$ to be %derived from the cumulative distribution
\begin{align}
\rmp_{1}(x_n;\rho_{n}) = \rho_n \rmq(x_n) + (1-\rho_n) \delta (x_n) \label{eq:cs_source}
\end{align}
%\begin{align}
%\rmP(x_n;\rho_{n}) = \rho_n \rmQ(x_n) + (1-\rho_n) \mone\set{x_n>0}, \label{eq:cs_source}
%\end{align}
for some distribution $\rmq(x_n)$. Then, $\bx$ models a sparse signal with non-uniform sparsity pattern whose non-zero entries are distributed with $\rmq(x_n)$. Consequently, by setting $u(v_n;c_n)=c_n\abs{v_n}$, the estimator reduces to the weighted $\ell_1$-norm minimization recovery scheme. For $\lambda_0=0$, the setup recovers the formerly studied noise-free case, e.g., \cite{tanaka2010optimal,khajehnejad2009weighted,scarlett2013compressed}, when $\lambda\downarrow 0$. 

In order to quantify the large-system performance of this setting, we define the weighted distortion as follows.
\begin{definition}[Weighted Distortion]
\label{def:dis}
Let $\bw_{N\times 1}$ enclose the coefficients $\set{w_n}$. The weighted distortion \ac{wrt} the distortion function $\sfd(\cdot;\cdot)$ for a given $\bw$ reads
\begin{align}
\sfD(\bx;\bhx|\bw) \coloneqq \frac{1}{N}{\sum_{n=1}^N w_n \E{\sfd(x_n;\hx_n)}}.\label{eq:Dw}
\end{align}
Moreover, the asymptotic weighted distortion is given by taking the limit $N\uparrow\infty$, i.e., $\sfD_{\bw} \coloneqq \lim_{N\uparrow\infty} \sfD(\bx;\bhx|\bw)$.

\end{definition}
The weighted distortion recovers various forms of recovery distortions. For instance, setting $w_n=1$ and $\sfd(x_n;\hx_n)=\abs{x_n - \hx_n}^2$, $\sfD_{\bw}$ determines the asymptotic \ac{mse}. Moreover, it evaluates the average error probability by setting $\sfd(x_n;\hx_n)=\mone\set{x_n = \hx_n}$ with $\mone\set{\cdot}$ being the indicator function. The main goal of this study is to derive the weighted distortion~in its generic form when $N$ grows large. %choosing $\set{w_n}$ proportional to $\set{\rho_n}$ and

\section{Asymptotic Performance}
\label{sec:main}
Invoking the replica method, $\sfD_{\bw}$ is derived in a closed form. The derivations are briefly sketched in Section~\ref{sec:large}. For the sake of compactness, we state the basic form of the result known as the ``\ac{rs} solution''. Our derivations are however in a general form enclosing ``\ac{rsb} solutions''. %Nevertheless, the \ac{rs} solution is believed to be sufficient for the regimes investigated in this paper. %We is given later in Section~\ref{sec:large}.% where we address~the~derivation of \ac{rsb} ans\"atze.% later in Section~\ref{sec:large}. %The detailed derivations for the extended version of the paper. We believe that for the specific scenarios which we consider here, the \ac{rs} ansatz gives a valid prediction.
\subsection{Asymptotic Weighted Distortion}
$\sfD_{\bw}$ in the large-system limit can be expressed in terms of an equivalent scalar system. For $j\in[J]$, we define the scalar estimator $\rmg_j^\dec(\cdot;c)$ which for given $c$ and $\theta$ reads
\begin{align}
\rmg_j^\dec(y;c)=\argmin_{v\in\setX}\frac{1}{2\theta} (y-v)^2 + u_j(v;c) \label{eq:dec_est}
\end{align}
$\rmg_j^\dec(y;c)$ represents a estimator which recovers a scalar from the single measurement $y$ using the one-dimensional form of \ac{map} formulation in \eqref{eq:map} with the weighting coefficient $c$ and estimation parameter $\theta$. In order to state the result, we moreover define the effective noise variance $\theta_0$ and the equivalent estimation parameter $\theta$ for some scalars $\chi$ and $\sfp$ as
%The function $\rmg_j(\cdot)$ is then defined as
%\begin{align}
%\rmg_j (c)\coloneqq \rmg_j^\dec(\xx_j+z^\rs;c)
%\end{align}
\begin{subequations}
\begin{align}
\theta &= \left[ \rmR_\mJ(-\frac{\chi}{\lambda})\right]^{-1} {\lambda}\\
\theta_0 &= \left[ \rmR_\mJ(-\frac{\chi}{\lambda})\right]^{-2} \frac{\partial}{\partial\chi} \left[(\lambda_0\chi-\lambda \sfp)\rmR_\mJ(-\frac{\chi}{\lambda})\right]
\end{align}
\end{subequations}
where $\rmR_\mJ(\cdot)$ denotes the $\rmR$-transform of $\rmp_\mJ(t)$ defined in the previous section. One should note that $\theta$ and $\theta_0$ are controlled by $\chi$ and $\sfp$ and are functions of the true estimation parameter $\lambda$, statistics of $\mA$ and the true noise variance $\lambda_0$.
\begin{proposition}%[\ac{rs} Ansatz]
\label{thm:1}
Let $z^\rs \sim \man\left(0,\theta_0\right)$, and for each $n\in[N]$, define the decoupled estimation $\rmg_n$ as
\begin{align}
\rmg_n \coloneqq \rmg^\dec_{j(n)}(x_n+z^\rs;c_n).
\end{align}
Then, under some assumptions\footnote{These assumptions are mainly replica continuity and the replica symmetry which are later introduced in Section \ref{sec:large}.}, $\sfD_{\bw}$ is given by
\begin{align}
\sfD_\bw &=\left\langle w_n \hspace*{.5mm} \E{\sfd \left( x_n;\rmg_n \right)} \right\rangle_{[N]}
\end{align}
where we define $\langle f(a_n) \rangle_\setN\coloneqq {\abs{\setN}}^{-1} \sum_{n\in\setN} f(a_n)$. The variables $\sfp$ and $\chi$ which determine $\theta$ and $\theta_0$ are moreover calculated from the fixed-point equations
\begin{subequations}
\begin{align}
\sfp&=\left\langle \E{ (\rmg_n-\xx_n)^2 } \right\rangle_{[N]}, \\
\frac{\theta_0}{\theta} \hspace*{.3mm} \chi &= \left\langle \E{ \left(\rmg_n-\xx_n\right) \hspace*{.3mm} z^\rs} \right\rangle_{[N]}.
\end{align}
\end{subequations}

\end{proposition}
\begin{prf}
The proof is briefly sketched in Section~\ref{sec:large}. The details of the proof, however, are skipped due to the page limitation.
\end{prf}
%By the same approach as in \cite{bereyhi2016rsb,bereyhi2016statistical}, we can justify the asymptotic decoupling property of this setup. To illustrate the property

\subsection{Asymmetric Decoupling Property}
\label{sec:main-B}
\begin{figure}
\begin{center}
\begin{tikzpicture}[auto, node distance=2.5cm,>=latex']
    \node [input, name=input] {};
    \node [sum, right of=input] (sum) {$+$};
    \node [input, above of=sum, node distance=.8cm] (noise) {};
    \node [block, right of=sum] (estimator) { $\mathrm{g}_{j(n)}(\cdot;c_n)$ };
    \node [output, right of=estimator, node distance=1.7cm] (output) {};

    \draw [draw,->] (input) -- node[pos=.1] {$\xx_n$} (sum) ;
    \draw [->] (sum) -- node[pos=.7] {$y_n$} (estimator);
    \draw [->] (estimator) -- node[pos=.9] [name=x_h] {$\hxx_n$}(output);
	\draw [draw,->] (noise) -- node[pos=.1] {$z^\dec$}(sum);
\end{tikzpicture}
\end{center}
	\caption{Asymmetric Decoupling Property: The decoupled systems are dependent on the index $n$ in general.\vspace*{-2mm}}
	\label{fig:1}
\end{figure}
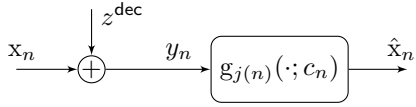

Proposition~\ref{thm:1} determines the asymptotic weighted distortion by averaging the scalar systems shown in Fig.~\ref{fig:1} over $n$ \ac{wrt} $\bw$. In fact by setting $\xx_n=x_n$ and $\hxx_n=\rmg_n$ in this diagram, one observes that $\sfD_{\bw}$ is the weighted average of input-output distortions. These scalar systems can be further shown to describe input-output marginal distributions. This observation states that the estimator exhibits the decoupling property in the large-system limit. To illustrate this property, let us denote the marginal joint distribution of $(\hx_n,x_n)$ with $\rmq_{N}(\hx_n,x_n)$ where the subscript indicates the dependency of the distribution on $N$. The asymptotic decoupling property mainly claims that as $N$ grows, $\rmq_{N}(\hx_n,x_n)$ converge to a deterministic distribution described by the input-output distribution of the scalar system in Fig.~\ref{fig:1}. % which describe a bank of decoupled single user systems. This means that for any $n\in[N]$, we have
%\begin{align}
%\lim_{N\uparrow\infty} \rmq_n^{N}(\hat{t},t) = \rmq_n(\hat{t},t) \label{eq:lim}
%\end{align}
%where $\rmq_n$ is a deterministic distribution. 
The previously studied forms of the property, e.g., \cite{bereyhi2016rsb,rangan2012asymptotic}, have considered identically distributed source entries, i.e., $\rmp_{j(n)}(\cdot;\rho_n)=\rmp(\cdot;\rho)$ and $u_{j(n)}(\cdot;c_n)=u(\cdot;c)$ for some constants $\rho$ and $c$. For this case, the limiting distribution is shown to be independent of $n$, and thus, the equivalent scalar systems are the same. The decoupled system derived in this paper, however, can vary from one index to another. We therefore refer to this form of decoupling as the ``asymmetric decoupling property'' which recovers the previous ``symmetric'' forms. The property is stated in the following. The proof follows the moment method and takes a similar path as in \cite{bereyhi2016statistical} with some modifications. It is however omitted for the sake~of~compactness.

\begin{decouple} %Using Proposition~\ref{thm:1}, we state an asymmetric form of the asymptotic decoupling property for the system specified in Section~\ref{sec:sys}. 
%The asymmetric decoupling property indicates that, Under some technical assumptions,
%\begin{inparaenum}
%\item the limit in \eqref{eq:lim} exists, and
%\item 
For $n\in[N]$, $(\hx_n,x_n)$ converges in distribution to the pair $(\hxx_n,\xx_n)$ in Fig.~\ref{fig:1} with $\rmg^\dec_{j}(\cdot;c_n)$ and $z^\dec$ being given in Proposition~\ref{thm:1}.
%in which $\xx_n \sim \rmp_{j(n)} \left( \cdot;\rho_n \right)$ and $\hxx_n=\rmg^\dec_{j(n)}(y_n;c_n)$ with $\rmg^\dec_{j}(\cdot;c_n)$ being defined in \eqref{eq:dec_est}. Here, $y_n$ reads
%\begin{align}
%y_n\coloneqq\xx_n+z^\dec \label{eq:dec_sys}
%\end{align}
%where $z^\dec$ is zero-mean Gaussian noise with~variance~$\theta_0$. Moreover, $\theta$ and $\theta_0$ are given in Proposition~\ref{thm:1}.
%\end{inparaenum}
\end{decouple}
%\begin{prf}
%
%\end{prf}

%The asymmetric decoupling indicates that the vector system in \eqref{eq:lin}, followed by the estimator $\bgg(\cdot)$ in \eqref{eq:map}, decouples in the large-system limit to a bank of effective scalar systems as in \eqref{eq:dec_sys}, followed by effective scalar estimators as in \eqref{eq:dec_est}.
%Fig.~\ref{fig:1} illustrates the decoupled setup for a given $n \in [N]$. Comparing this property to scenarios with symmetry at the source and estimator, one observes that the decoupled estimators in asymmetric setups are indexed by $n$. The effective noise powers are however the same in all the decoupled setups. This observation is consistent with the intuition which indicates that despite the asymmetry at the source and estimator, the overall impact of the interference at each observation is symmetric.
\section{Applications of the Main Results}
\label{sec:app}
%As it was discussed, the setup described in Section~\ref{sec:sys} recovers several formerly studied systems. Therefore, 
The asymptotic results presented in Section~\ref{sec:main} can be employed to investigate various estimation problems. In the sequel, we give some examples in compressive sensing.% of these applications. % in compressive sensing and multiuser communications.
%The generality of the setup, moreover, enables us to study scenarios which have not been addressed in the literature. 
%To give examples, we employ the results to study some problems in compressive sensing and multiuser communications. Further illustrations however are skipped due to the page limitation.% and left for the extended version of the manuscript. %In this section, we consider some applications of the results given in Section~\ref{sec:main}.
%\vspace*{-2mm}
\subsection{Recovery of Non-uniform Sparse Signals}
Stochastic signals with non-uniform sparsity patterns are described by our setting when $J=1$ and the signal entries $x_n$ are distributed as in \eqref{eq:cs_source}. Several recovery schemes, some of which have not been addressed in the literature, can then be investigated by choosing corresponding utility functions. A trivial approach is to let the utility function be %a weighted $\ell_p$-norm which means that
\begin{align}
u(\bv;\bc)=\sum_{n=1}^N c_n\abs{v_n}^p.
\end{align}
Using Proposition~\ref{thm:1}, the large-system performance of these recovery schemes can be studied \ac{wrt} various forms of distortions. Moreover, the optimal choices for $\set{c_n}$ can be found in terms of the priors $\set{\rho_n}$, such that the average distortion is minimized. This investigation widens the scope of analyses in~\cite{tanaka2010optimal} to noisy scenarios and various recovery schemes. Moreover, it enables us to extend the recent study in \cite{zheng2017does} to cases with prior information on the sparsity pattern. To discuss further the application of the results in recovery of non-uniform sparse signals, we consider the following example.

\begin{figure}[t]
\hspace*{-.2cm}  
%\centering
\resizebox{1\linewidth}{!}{
\pstool[width=.35\linewidth]{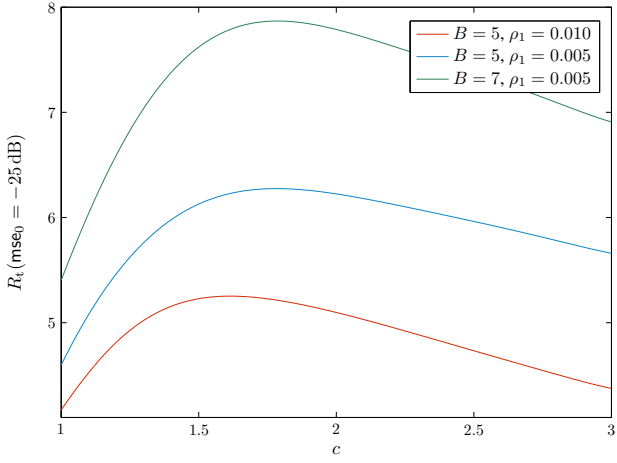}{
\psfrag{Rth}[c][t][0.26]{$R_{\rmt}(\mse_0=-25 \hspace*{.5mm} \rm dB)$}
\psfrag{c}[c][c][0.26]{$c$}
\psfrag{B=5-RR=10-DDDDD}[l][l][0.25]{$B=5$, $\rho_1=0.010$}
\psfrag{B=5-RR=05-DDDDD}[l][l][0.25]{$B=5$, $\rho_1=0.005$}
\psfrag{B=7-RR=05-DDDDD}[l][l][0.25]{$B=7$, $\rho_1=0.005$}
%y-axis

\psfrag{4}[r][c][0.22]{$4$}
\psfrag{5}[r][c][0.22]{$5$}
\psfrag{6}[r][c][0.22]{$6$}
\psfrag{7}[r][c][0.22]{$7$}
\psfrag{8}[r][c][0.22]{$8$}

%
%%x-axis
\psfrag{1}[c][b][0.22]{$1$}
\psfrag{2}[c][b][0.22]{$2$}
\psfrag{3}[c][b][0.22]{$3$}
\psfrag{1.5}[c][b][0.22]{$1.5$}
\psfrag{2.5}[c][b][0.22]{$2.5$}

}}
\caption{$R_{\rmt}(\mse_0)$ at $\mse_0=-25$ dB versus the norm-weight of the more sparse block in Example \ref{ex:1}. As the figure shows, either the growth in the size of the block or reduction in its sparsity increases the degradation caused by uniform recovery, i.e., $c=1$.\vspace*{-2mm}}
\label{fig:2}
\end{figure}

%Moreover, our general replica analysis, which is presented later in Section~\ref{sec:large}, enables us to investigate the cases in which the \ac{rs} ansatz is not valid by determining~the~\ac{rsb} ans\"atze as well. 
\begin{example}
\label{ex:1}
Assume that $\bx$ is a sparse-Gaussian signal with a non-uniform sparsity pattern, i.e., $J=1$ and the distribution of $x_n$ for $n\in[N]$ are given by $\rmp_1(x_n;\rho_{n})$ in \eqref{eq:cs_source} %
%\begin{align}
%\rmp(x_n;\rho_{n}) = \rho_n \phi \left( x_n \right) + \left( 1-\rho_n \right) \delta \left( x_n \right) \label{eq:cs_source}
%\end{align}
with $\rmq(x_n)$ being the zero-mean and unit-variance Gaussian distribution. To recover the signal, we employ the weighted zero-norm recovery scheme which is given by setting $u(v_n;c_n)=c_n \mone \set{v_n\neq 0}$ in \eqref{eq:map}. Proposition~\ref{thm:1}~enables us to investigate the recovery performance in this case and also evaluate the optimal choice of $\set{c_n}$ in terms of $\set{\rho_n}$. For the sake of simplicity, consider the scenario in which $\mA$~is~an \ac{iid} matrix whose entries are zero-mean with variance $1/K$. In this case, $\rmp_\mJ$ follows the Marcenko-Pastur law \cite{muller2013applications}, and thus, $\rmR_\mJ(\omega) = {\alpha}{(\alpha-\omega)^{-1}}$ which implies $\theta=\lambda+ \alpha^{-1} \chi$ and $\theta_0=\lambda+ \alpha^{-1} \sfp$. Moreover, %$\rmg^\dec(\cdot;c)$ reads
\begin{align}
\rmg^\dec(y_n;c_n)=
\begin{cases}
y_n &\abs{y_n}> t_n\\
0 &\abs{y_n}\leq t_n
\end{cases} \label{eq:zero_norm}
\end{align}
where $t_n \coloneqq \sqrt{2\theta c_n}$. Consequently, the asymptotic distortion \ac{wrt} some given distortion function and $\bw$ is determined by Proposition~\ref{thm:1}. As \eqref{eq:zero_norm} shows, weighted zero-norm recovery decouples asymptotically into a set of hard thresholding operators whose threshold levels depend on weights $c_n$. By setting $c_n=1$ and $\rho_n=\rho$ for all $n\in [N]$, the decoupled setups reduce to the symmetric setups reported in \cite{rangan2012asymptotic,bereyhi2016rsb}. 

To investigate the performance of weighted zero-norm recovery numerically, we consider the configuration in which %of the setup $\set{\rho_n}$ is given by
%\begin{align}
%\rho_n=
%\begin{cases}
%\rho_0 &n\in[{N}/{3}],\\
%\rho_1 &n\in[{N}/{3}+1:{2N}/{3}],\\
%\rho_2 &n\in[{2N}/{3}+1:{N}]
%\end{cases}
%\end{align}
%for some $\rho_0$, $\rho_1$ and $\rho_2$ in $[0,1]$. Correspondingly, we set $\set{c_n}$
%\begin{align}
%c_n=
%\begin{cases}
%1 &n\in[{N}/{3}],\\
%c_1 &n\in[{N}/{3}+1:{2N}/{3}],\\
%c_2 &n\in[{2N}/{3}+1:{N}]
%\end{cases}
%\end{align}
\begin{align}
\rho_n=
\begin{cases}
\rho_0 &n\in[{N}/{B}],\\
\rho_1 &n\in[{N}/{B}+1:N],
\end{cases}
\end{align}
for some $\rho_0,\rho_1 \in [0,1]$ and some integer $B$ being a divisor of $N$. Here, $[M:N]$ denotes $\set{M,\ldots,N}$. Moreover, we set %$\set{c_n}$ to be
\begin{align}
c_n=
\begin{cases}
1 &n\in[{N}/{B}],\\
c &n\in[{N}/{B}+1:N],
\end{cases}
\end{align}
for some $c$. We denote the asymptotic average \ac{mse} by $\mse \coloneqq \lim_{N\uparrow\infty} \E{\norm{\bx-\bhx}^2}/N$. Moreover, for a given $\mse_0$, we define the threshold compression rate $R_{\rmt}(\mse_0)$ to be the maximum possible inverse load factor $\alpha^{-1}=N/K$ which results in $\mse\leq \mse_0$. %asymptotic average \ac{mse} of the system defined as
Fig.~\ref{fig:2} shows the threshold compression rate as a function of $c$ for $\mse_0=-25$ dB.  The curves have been plotted for $\rho_0=0.1$ considering various choices of $\rho_1$ and $B$. The noise power is set to be $\lambda_0=0.01$ and $\lambda$ is tuned such that the \ac{mse} is minimized at each load factor. As the figure shows, the optimal choice of $c$ can significantly increase the threshold compression rate. The curves moreover indicate that as $B$ grows or $\rho_1$ reduces the gap between the optimal $R_{\rmt}(\mse_0)$, maximized over $c$, and the threshold compression rate at $c=1$ increases. This observation is due to the fact that the growth in $B$ or the reduction in $\rho_1$ imposes more asymmetry into the setting, and therefore, increases the loss caused by uniform recovery, i.e., $c=1$.
%a negligible gain for small rates of compression. The gain however starts to increase as the compression rate grows.
%considering equal weights for all entries %the asymptotic distortion for a 
\end{example}
%\begin{remark}
%For the specific setups whose corresponding \ac{mse} curves have been sketched in Fig.~\ref{fig:2}, one could also evaluate \ac{mse} by splitting the setup into a 
%\end{remark}

%The curves report a significant gain at optimal choice of $c$ compared to the case with $c=1$. The gain increases as $B$ grows or $\rho_1$ reduces.
\vspace*{-2mm}
\subsection{Non-uniform Sparsity with Multiple Prior States}
The non-uniform sparsity model can be extended to signals with multiple prior states by considering $J>1$. In this case, % the source distribution to
\begin{align}
\rmp_{j(n)}(x_n;\rho_{n}) = \rho_n \rmq_{j(n)}(x_n) + (1-\rho_n) \delta(x_n)
\end{align}
represents a signal with non-uniform sparsity pattern whose non-zero entries are taken from multiple possible prior distributions. This model describes a scenario in which multiple uncorrelated non-uniform sparse signals are simultaneously measured, e.g., a network of independent sensors with different prior distributions. An efficient approach for signal recovery in this case is to set $u_{j(n)}(v_n;c_n)= f_{j(n)}(v_n) + c_n \mone\set{v_n\neq 0}$ for some $f_j(v)$. Similar to the case with single state sources, the optimal choice of $\set{c_n}$ as well as the asymptotic distortion is determined using Proposition~\ref{thm:1}.

\section{Large-System Analysis}
\label{sec:large}
In this section, we briefly sketch the derivations based on the replica method. % ansatz which includes the \ac{rs} as well as \ac{rsb} ans\"atze. %The derivations %are based on the replica method, developed for the analysis of spin glasses \cite{edwards1975theory}, and 
%take the following two steps.
%\begin{enumerate}[label=(\alph*)]
%\item We determine the asymptotic weighted distortion in terms of the free energy of a corresponding spin glass.
%\item The replica method is employed to evaluate the~free~energy function.
%\end{enumerate}
%For sake of brevity, throughout this section, we assume $\setX$ to be discrete. The analysis, however, is in full generality and straightforwardly extends to continuous choices of $\setX$. 
%\subsection{Corresponding Spin Glass}
%A thermodynamic system is a significantly large multi-body system. The state of a thermodynamic system is denoted by the microstate vector whose entries represent the state of system's particles. In spin glasses, the microstate takes random level of energy. 
Consider %the Hamiltonian %$\mae(\cdot|\cdot)$ which for given realizations of $\mA$ and $\by$ reads
$\mae(\bv)=\norm{\by-\mA\bv}^2/2 \lambda + u(\bv;\bc)$, and define
%In the context of statistical mechanics, \eqref{eq:ham} denotes the Hamiltonian of a spin~glass~with quenched variables $\by$ and $\mA$. For this Hamiltonian the modified partition function is given by%whose microstates, i.e., $\bv\in\setX^N$, are conditionally distributed with
%\begin{align}
$\maz(\beta,h) =  \sum_{\bv} \exp\{-\beta\mae(\bv) +hN\sfD(\bx;\bv|\bw)\}$.
%\end{align}
%and the free energy of the spin glass reads
%\begin{align}
%\maf(\beta,h) =\frac{1}{N} \E{ \log \maz(\beta,h)}.
%\end{align}
%
%%  and define the partition function $\maz(\beta,h)$ to be
%\begin{align}
%\maz(\beta,h) =\sum_{\bv} e^{-\beta\mae(\bv|\bs,\mH)+hn\sfM^{\setW}_f(\bv;n)}.
%\end{align}
%\subsection{Evaluation of the Asymptotic Marginal}
%\label{ssec:asy_marg}
%The above spin glass is denoted as the corresponding spin glass to the sampling system, and is shown to determine the asymptotics of the estimator given in \eqref{eq:map} \cite{bereyhi2016statistical}. Considering the asymptotic weighted distortion in Definition \ref{def:dis}, 
One can then employ large deviation arguments and write% of the sampling system is determined from the free energy function. To illustrate the latter statement, define the macroscopic parameter $\sfD(\beta)$ as
\begin{align}
\sfD_\bw =\lim_{N\uparrow\infty} \lim_{\beta\uparrow\infty} \frac{\partial}{\partial h} \maf(\beta,h)|_{h=0}. \label{eq:10}
\end{align}
where $\maf(\beta,h) = \E{ \log \maz(\beta,h)}/N$. %
%\eqref{eq:10} evaluates the asymptotic distortion in terms of the~free~energy. The relation enables us to invoke mathematical~tools~developed in statistical mechanics and overcome the non-trivial task of determining the asymptotic distortion. Determining the free energy confronts us with the non-traceable task of evaluating a logarithmic expectation.~The~task turns out to be non-feasible when the function under~the~logarithm is a sum of exponential terms.
%
%\subsection*{The Replica Method}
As evaluating a logarithmic expectation is not a trivial task, we invoke the replica method. %The replica method is a non-rigorous but effective mathematical tool which tries to come over this hard task by means of some large-system tricks. 
The main idea comes from the Riesz equality \cite{riesz1930valeurs} which states $\E{\log \xx} = \lim_{m\downarrow 0} \log \E{\xx^m}/m$. Using this equality, $\sfD_\bw$ is determined in terms of the $m$th moment of $\maz(\beta,h)$. Nevertheless, the moments need to be determined for real values of $m$ which is still challenging. This challenge is addressed by assuming ``replica continuity'' which means that $\E{\maz^{m}(\beta,h)}$ analytically continues from $m\in\setZ^+$ to $m\in\setR^+$. %Assuming replica continuity, $\maf(\beta,h)$ is analytically determined, and by substituting into \eqref{eq:10},  
After calculating the moments, $\sfD_\bw$ is given by
\begin{align}
\sfD_\bw=\lim_{\beta\uparrow\infty} \lim_{m\downarrow0} \sum_{\bvv} \left\langle\E{w_n \sfd(\bvv;\bxx_n)\rmp_{n}^\beta(\bvv|\bxx_n)}\right\rangle_{[N]}, \label{eq:gen}
\end{align}
%Consider the nonlinear \ac{lse} precoder in Section \ref{sec:sys}, and define $\bvv_{m\times1}$ to be a random vector over $\setX^m$ with the~distribution~
for $\bvv\in\setX^m$, where $\bxx_n$ is an $m\times1$ vector with all the entries being $x_n$ and $\sfd(\bvv;\bxx_n)\coloneqq \sum_{a=1}^m \sfd(\vv_a;x_n)$; moreover,% and %$\rmp_{n}^\beta (\bvv;\mQ|\bx)$ reads
\begin{align}
\rmp_{n}^\beta (\bvv|\bxx_n)&=\dfrac{e^{-\beta\left[(\bvv-\bxx_n)^\her \mR (\bvv-\bxx_n)+ u_{j(n)}(\bvv;c_n)\right] }}{\sum_{\bvv} e^{-\beta\left[(\bvv-\bxx_n)^\her \mR (\bvv-\bxx_n)+ u_{j(n)}(\bvv;c_n)\right] }}. 
\end{align}
with $u_j(\bvv;c_n)\coloneqq \sum_{a=1}^m u_j(\vv_a;c_n)$, and $\mR\coloneqq \mT \rmR_{\mJ} (-\beta \mT \mQ)$ for $\mT\hspace*{-.7mm}=\hspace*{-.7mm}\dfrac{1}{2\lambda} (\mI_m \hspace*{-.7mm}-\hspace*{-.7mm}\beta\dfrac{\lambda_0}{\lambda} \mone_m)$ and some $\mQ_{m\times m}$ which satisfies
\begin{align}
\mQ&=\sum_{\bvv} \left\langle\E{ \rmp_{n}^\beta (\bvv|\bxx_n) (\bvv-\bxx_n) (\bvv-\bxx_n)^\her}\right\rangle_{[N]}. \label{eq:Q}
\end{align}
In \eqref{eq:gen}, the general replica solution is given.~The~explicit~determination of $\sfD_\bw$, however, needs ${\mQ}$ to be found such that \eqref{eq:Q} is fulfilled. %Due to both the complexity and the analyticity issues, the replica method suggests to consider some initial structures on ${\mQ}$. These structures come from the physical properties of spin glasses. 
%The common trick in statistical mechanics is 
To do so, we need to suppose a structure for $\mQ$. The basic structure is given by \ac{rs} as ${\mQ}= {\chi}{\beta}^{-1} \mI_m + \sfp \mone_m$ for some $\chi$ and $\sfp$. By substituting $\mQ$ in \eqref{eq:gen}, Proposition~\ref{thm:1} is concluded after some lines of derivations. The \ac{rsb} solutions are further derived by extending the \ac{rs} structure to
\begin{align}
\mQ= \frac{\chi}{\beta} \mI_m + \sum_{\kappa=1}^b \sfc_\kappa\hspace*{.5mm} \mI_{\frac{m\beta}{\mu_\kappa}} \otimes \mone_{\frac{\mu_\kappa}{\beta}} +  \sfp \mone_m, \label{eq:rsb}
\end{align}
for some integer $b$. The derivations under \ac{rsb} follow \cite[Appendix D]{bereyhi2016statistical} and are omitted due to the page limitation. 
\section{Conclusion}
In this paper, we have studied the asymptotic performance of  a class of \ac{map}-based signal recovery schemes when prior information is available for reconstruction. Our analysis has demonstrated an asymmetric version of the decoupling principle for these estimators which generalizes the formerly studied forms of \ac{map} decoupling \cite{rangan2012asymptotic,bereyhi2016rsb}. Invoking the results, we have investigated the performance of weighted zero-norm minimization for recovery of a signal with non-uniform sparsity pattern. The results of this paper can be further employed to study various problems. A particular example in compressive sensing is to extend the scope of investigations in \cite{zheng2017does} to signals with non-uniform sparsity patterns and study the impact of replacing the $\ell_1$-norm with an $\ell_p$-norm in the weighted norm minimization scheme for $0\leq p\leq1$. Currently, the work in this direction has been started.

\newpage

\bibliographystyle{IEEEbib}
\bibliography{ref}

\begin{acronym}
\acro{mse}[MSE]{Mean Square Error}
\acro{mimo}[MIMO]{Multiple-Input Multiple-Output}
\acro{csi}[CSI]{Channel State Information}
\acro{awgn}[AWGN]{Additive White Gaussian Noise}
\acro{iid}[i.i.d.]{independent and identically distributed}
\acro{ut}[UT]{User Terminal}
\acro{bs}[BS]{Base Station}
\acro{tas}[TAS]{Transmit Antenna Selection}
\acro{lse}[LSE]{Least Square Error}
\acro{rhs}[r.h.s.]{right hand side}
\acro{lhs}[l.h.s.]{left hand side}
\acro{wrt}[w.r.t.]{with respect to}
\acro{rs}[RS]{Replica Symmetry}
\acro{rsb}[RSB]{Replica Symmetry Breaking}
\acro{papr}[PAPR]{Peak-to-Average Power Ratio}
\acro{rzf}[RZF]{Regularized Zero Forcing}
\acro{snr}[SNR]{Signal-to-Noise Ratio}
\acro{rf}[RF]{Radio Frequency}
\acro{map}[MAP]{Maximum-A-Posteriori}
\acro{pmf}[PMF]{Probability Mass Function}
\acro{pdf}[PDF]{Probability Density Function}
\acro{cdf}[CDF]{Cumulative Distribution Function}
\end{acronym}

\end{document}

%% file: commands.tex
\newcommand{\setX}{\mathbbmss{X}}

\newcommand{\setR}{\mathbbmss{R}}

\newcommand{\setZ}{\mathbbmss{Z}}
\newcommand{\setB}{\mathbbmss{B}}

\newcommand{\setN}{\mathbbmss{N}}

\DeclareMathOperator*{\argmin}{argmin}

\newcommand{\rmp}{\mathrm{p}}

\newcommand{\rmq}{\mathrm{q}}

\newcommand{\rmR}{\mathrm{R}}
\newcommand{\rmg}{\mathrm{g}}
\newcommand{\rmG}{\mathrm{G}}

\newcommand{\dec}{\mathsf{dec}}

\newcommand{\her}{\mathsf{H}}
\newcommand{\sfc}{\mathsf{c}}

\newcommand{\sfD}{\mathsf{D}}

\newcommand{\sfd}{\mathsf{d}}

\newcommand{\sfp}{\mathsf{p}}

\newcommand{\maf}{\mathcal{F}}

\newcommand{\mae}{\mathcal{E}}
\newcommand{\maz}{\mathcal{Z}}

\newcommand{\man}{\mathcal{N}}

\newcommand{\bxx}{\mathbf{x}}

\newcommand{\bvv}{\mathbf{v}}

\newcommand{\bx}{{\boldsymbol{x}}}
\newcommand{\hx}{{\hat{x}}}

\newcommand{\vv}{\mathrm{v}}
\newcommand{\hxx}{\hat{\mathrm{x}}}
\newcommand{\xx}{\mathrm{x}}

\newcommand{\bhx}{{\boldsymbol{\hat{x}}}}
\newcommand{\set}[1]{\left\lbrace#1\right\rbrace}

\newcommand{\bz}{{\boldsymbol{z}}}

\newcommand{\bv}{{\boldsymbol{v}}}
\newcommand{\bc}{{\boldsymbol{c}}}
\newcommand{\bw}{{\boldsymbol{w}}}

\newcommand{\by}{{\boldsymbol{y}}}

\newcommand{\trp}{\mathsf{T}}

\newcommand{\mA}{\mathbf{A}}
\newcommand{\mR}{\mathbf{R}}

\newcommand{\mI}{\mathbf{I}}
\newcommand{\mone}{\mathbf{1}}
\newcommand{\mJ}{\mathbf{J}}

\newcommand{\mQ}{\mathbf{Q}}

\newcommand{\mU}{\mathbf{U}}
\newcommand{\mD}{\mathbf{D}}

\newcommand{\mT}{\mathbf{T}}

\newcommand{\rmt}{\mathrm{t}}

\newcommand{\E}[1]{\mathbb{E} \left\lbrace #1 \right\rbrace }

\newcommand{\mse}{\mathsf{mse}}

\newcommand{\rs}{{\mathsf{dec}}}

\newcommand{\norm}[1]{\lVert #1 \rVert}

\newcommand{\abs}[1]{\lvert #1 \rvert}

\newtheoremstyle{mystyle}%                % Name
  {}%                                     % Space above
  {}%                                     % Space below
  {}%                                     % Body font
  {}%                                     % Indent amount
  {\bfseries}%                            % Theorem head font
  {:}%                                     % Punctuation after theorem head
  { }%                                    % Space after theorem head, ' ', or \newline
  {}%                                     % Theorem head spec (can be left empty, meaning `normal')

\theoremstyle{mystyle}

\newtheorem{definition}{Definition}
\newtheorem{proposition}{Proposition}

\newtheorem{example}{Example}
\newtheorem*{prf}{Proof}

\newtheorem*{decouple}{Asymmetric Decoupling}
\newcounter{bar}